\documentclass[preprint,aps,nofootinbib]{revtex4}
\usepackage{graphicx}
\usepackage{epstopdf}
\usepackage{subfigure}
\usepackage{textcomp}

\usepackage{dsfont}
\usepackage{amsfonts}
\usepackage{amsmath}
\usepackage{amssymb}
\def\bkR{{\rm I\kern-.17em R}}
\def\bkC{{\rm \kern.24em \vrule width.05em height1.4ex depth-.05ex \kern-.26em C}}

\def\be{\beta}

\def\frac#1#2{{\textstyle{{#1}\over {#2}}}}

\def\lsim{\mathrel{\rlap{\lower4pt\hbox{\hskip1pt$\sim$}}
    \raise1pt\hbox{$<$}}}
\def\gsim{\mathrel{\rlap{\lower4pt\hbox{\hskip1pt$\sim$}}
    \raise1pt\hbox{$>$}}}
\def\sqr#1#2{{\vcenter{\vbox{\hrule height.#2pt
         \hbox{\vrule width.#2pt height#1pt \kern#1pt
         \vrule width.#2pt}
         \hrule height.#2pt}}}}

\def\laq{\raise 0.4 ex \hbox{$<$}\kern -0.8 em\lower 0.62 ex\hbox{$\sim$}}
\def\gaq{\raise 0.4 ex \hbox{$>$}\kern -0.7 em\lower 0.62 ex\hbox{$\sim$}}

\def\be{\begin{equation}}
\def\ee{\end{equation}}
\def\beqa{\begin{eqnarray}}
\def\eeqa{\end{eqnarray}}

\def\dalemb#1#2{{\vbox{\hrule height.#2pt
        \hbox{\vrule width.#2pt height#1pt \kern#1pt \vrule width.#2pt}
        \hrule height.#2pt}}}

\def\dalemb#1#2{{\vbox{\hrule height.#2pt
        \hbox{\vrule width.#2pt height#1pt \kern#1pt \vrule width.#2pt}
        \hrule height.#2pt}}}

\def\gtorder{\mathrel{\raise.3ex\hbox{$>$}\mkern-14mu
             \lower0.6ex\hbox{$\sim$}}}
\def\ltorder{\mathrel{\raise.3ex\hbox{$<$}\mkern-14mu
             \lower0.6ex\hbox{$\sim$}}}

\begin{document}

\rightline{September 2011}

\title{Phase-Space Noncommutativity and the Dirac Equation}

\author{Orfeu Bertolami\footnote{Also at Instituto de Plasmas e Fus\~ao Nuclear,
Instituto Superior T\'ecnico, Avenida Rovisco Pais 1, 1049-001 Lisboa, Portugal. E-mail: orfeu.bertolami@fc.up.pt}}

\vskip 0.3cm

\affiliation{Departamento de F\'\i sica e Astronomia, Faculdade de Ci\^encias da Universidade do Porto \\
Rua do Campo Alegre 687, 4169-007 Porto, Portugal}

\author{Raquel Queiroz\footnote{E-mail: rpa.queiroz@gmail.com}}

\vskip 0.3cm

\affiliation{Instituto de Plasmas e Fus\~ao Nuclear, Instituto Superior T\'ecnico \\
Avenida Rovisco Pais 1, 1049-001 Lisboa, Portugal
\vspace{2cm}
}


\vskip 0.5cm

\begin{abstract}

\vskip 0.5cm

{
We consider full phase-space noncommutativity in the Dirac equation, and find that in order to preserve gauge invariance, configuration space noncommutativity must be dropped. The resulting space structure gives rise to a constant magnetic field background and this effect is explicitly seen on the spectrum of the hydrogen atom. Computing this spectrum we find a bound on the momentum noncommutative parameter $\eta$, $\sqrt{\eta}\lsim2.26\mu eV/c$.
}

\end{abstract}

\maketitle

\section{Introduction}

The study of noncommutative (NC) spaces and their implications for quantum field theories, noncommutative quantum field theories, is an extremely active area of research (see e.g. Refs. \cite{DS} for reviews).

A noncommutative geometry is defined by the introduction of an antisymmetric constant matrix, $\theta$, of dimensions of (length)$^2$, given by
\begin{align}
[x_i,x_j]=i\theta_{ij},\label{first}
\end{align}
where $i,j$ stand for spatial indices.
Noncommutativity is the central mathematical concept describing uncertainty, so an immediate consequence of this geometry is that
\begin{align}
\Delta x_i \Delta x_j \ge {1\over 2}|\theta_{ij}|,
\end{align}
which introduces a space cell scale, $l_\text{NC}=\sqrt{\theta}$ \cite{DS}. From Eq. (\ref{first}), it is clear that noncommutativity leads to non-local features and the breaking of Lorentz symmetry \cite{Carroll} (see however Ref. \cite{BGuisado2}), which may be some of the ingredients of quantum gravity. In fact, noncommutativity is encountered in string theory induced by a non-trivial NS B-field background \cite{CS} .  

Noncommutative extensions of quantum mechanics have been the focus of active research \cite{Ho,Nair,Duval,Horvathy,4,5,OB1,7,8,OB3}, and the extensions which include noncommutativity in momenta have quite rich implications for quantum cosmology \cite{9} and black holes \cite{10,11,12}. 

In this work we study the phenomenology of a noncommutative version of the Dirac Hamiltonian, in which noncommutativity is considered in both coordinate and momentum spaces. The configuration space noncommutative features of this system were first addressed in Ref. \cite{Gitman}. The hydrogen energy spectrum is then used to constrain the momentum noncommutative parameter.

This article is organized as follows. In the first section we introduce the NC algebra and its mathematical background. In section III we compute the appropriate corrections to the Dirac Hamiltonian and explore the dynamics of a particle described by it. In section IV we find a bound on the momentum noncommutative parameter using the spectrum of the hydrogen atom. Finally, section V contains our conclusions.

\section{Phase-space noncommutativity}

Consider a noncommutative algebra satisfying the commutation relations:
\begin{align}
[x_i,x_j]=i\theta_{ij}, && [p_i,p_j]=i\eta_{ij}, && [x_i,p_j]=i\hbar\delta_{ij}, \label{noncom}
\end{align}
where $\theta_{ij}$ and $\eta_{ij}$ are antisymmetric constant matrices with dimensions of (length)$^2$ and (momentum)$^2$, respectively. 

The product of functions of noncommuting variables, i.e., the fields in the noncommuting space, satisfy the generalized Moyal product \cite{OB3}, which for a vanishing $\eta_{ij}$ corresponds to the usual Moyal product:
\begin{align}
f(x)\star g(x)\equiv f(x)\exp\left\{{i\over 2}\overleftarrow{\partial_i}\theta_{ij}\overrightarrow{\partial_j}\right\}g(x)\approx f(x)g(x)+{1\over 2}\theta_{ij}\partial_if(x)\partial_jg(x)+\mathcal{O}(\theta^2).\label{moyal}
\end{align}

The NC algebra can be mapped into the commutative Heisenberg-Weyl algebra, through a non-canonical linear transformation $(x,p)\mapsto(x',p')$ such that the new variables satisfy:
\begin{align}
[x'_i,x'_j]=0, && [p'_i,p'_j]=0, && [x'_i,p'_j]=i\hbar\delta_{ij}.
\end{align}
We can choose for instance the following map:
\begin{align}
x_i=\left(x'_i-{\theta_{ij}\over 2\hbar}p'_j\right), && p_i=\left(p'_i+{\eta_{ij}\over 2\hbar}x'_j\right),\label{map}
\end{align}
so \eqref{noncom} becomes\footnote{The commutator $[x,p]$ is modified for consistency, and in particular it is no longer diagonal. The non-diagonal terms cancel if one sets one of the parameters to zero.}:
\begin{align}
[x_i,x_j]=i\theta_{ij}, && [p_i,p_j]=i\eta_{ij}, && [x_i,p_j]=i\hbar\left(\delta_{ij}+{\theta_{ia}\eta_{ja}\over 4\hbar^2}\right).
\end{align}

Although this map is not unique, it has been proved that physical predictions, i.e. expectation values, transition rates, etc., are independent of the chosen map \cite{OB3}. Since the fields do not depend on derivatives, we will only consider functions of $x$. Hence, in the following discussion, the dependence on the derivatives will be mapped using Eq. \eqref{map}. 

\section{Noncommutative Dirac Hamiltonian}

The Dirac Hamiltonian is given by:
\begin{align}
H=(c\boldsymbol{\alpha}\cdot(\mathbf{p}-e\mathbf{A})+\beta mc^2+e\Phi), \label{DH}
\end{align}
where the momentum $p$ is given by $p_i\equiv -i\hbar\partial_i$ and the matrices $\alpha_i$ and $\beta$ satisfy the anti-commutation relations
\begin{align}
\{\alpha_i,\alpha_j\}=2\delta_{ij}, && \{\alpha_i,\beta\}=0, && \alpha_i^2=\beta^2=\mathds{1}
\end{align}
and take the explicit form \begin{equation}
\alpha_i=
\begin{bmatrix}
0 & \sigma_i \\
\sigma_i & 0
\end{bmatrix}, \hspace{30pt}
\beta=
\begin{bmatrix}
1 & 0 \\
0 & -1
\end{bmatrix} ,
\end{equation}
which correspond to $\gamma$-matrices when the Dirac Hamiltonian is represented in a covariant form.
The coupling with the electromagnetic potential, $A^\mu=({\Phi\over c}, \mathbf{A})$, is minimally introduced with a charge $e$ using the gauge invariance of the equation, and we can define the canonical momentum as $\boldsymbol{\pi}\equiv \mathbf{p}-e\mathbf{A}$.

To introduce noncommutativity into the Dirac Hamiltonian the map in Eq. \eqref{map} is used after substituting the product of functions by the Moyal product Eq. \eqref{moyal}. For simplicity, $\theta_{ij}$ and $\eta_{ij}$ matrices will be written as $\theta_{ij}=\theta\epsilon_{ij}$ and $\eta_{ij}=\eta\epsilon_{ij}$, where $\epsilon_{ij}$ is an antisymmetric unitary\footnote{Unitary in this context means that ${1\over 2}\varepsilon_{ijk}\epsilon_{ij}$ is a normalized vector $e_k$, $\varepsilon_{ijk}$ being the Levi-Civita symbol. } matrix. Note that this is not uniquely defined since $i,j=1,2,3$.

To first order in $\theta_{ab}$:
\begin{align}
H(x,p)\star\Psi(x)&=H(x',p)\Psi(x')+{i\over 2}\theta_{ab}\partial_a\left(c\alpha_i(p_i-eA_i(x'))+\beta mc^2+e\Phi(x')\right)\partial_b\Psi(x') \nonumber
\\
&=H(x',p)\Psi(x')-{ie\over 2}\theta_{ab}\partial_a(c\alpha_iA_i(x')-\Phi(x'))\partial_b\Psi(x').
\end{align}
Using Eq. \eqref{map} we find the Hamiltonian dependence on commuting coordinates to first order in $\theta_{ab}$ and $\eta_{ab}$:
\begin{align}
H(x',p')\Psi(x')&=\left[c\alpha_i(p'_i+{1\over 2\hbar}\eta_{ij}x'_j-eA_i(x'))+\beta mc^2+e\Phi+{ie\over 2}\theta_{ab}\partial_a(c\alpha_iA_i(x')-\Phi(x'))\partial_b\right]\Psi(x').
\end{align}
For convenience we drop the primes and rewrite the the new Hamiltonian in a more compact form\footnote{$\theta_k\equiv{1\over 2}\varepsilon_{kij}\theta_{ij}$ and $\eta_k\equiv{1\over 2}\varepsilon_{kij}\eta_{ij}$.}:
\begin{align}
H_{NC}\equiv c\boldsymbol{\alpha}\cdot(\mathbf{p}-e\mathbf{A})+\beta mc^2 + e\Phi-{e\over 2\hbar}[\nabla(c\boldsymbol{\alpha}\cdot\mathbf{A}-\Phi)\times\mathbf{p}]\cdot\boldsymbol{\theta}+{c\over 2\hbar}[\boldsymbol{\alpha}\times\mathbf{r}]\cdot\boldsymbol{\eta}.\label{HAM}
\end{align}

\subsection{Dynamics of a charged fermion}

For a particle subject to a constant electromagnetic field, we can write the potential up to a constant,
$\mathbf{A}={1\over 2}\mathbf{B}\times\mathbf{r}$, $\Phi=\mathbf{r}\cdot[\mathbf{E}+{1\over 2}\mathbf{v}\times\mathbf{B}]$, and using these identities we obtain  the equations of motion for a charged particle:
$
\mathbf{v}=c\boldsymbol{\alpha}\label{velocidade}$, $
\mathbf{F}=e[\mathbf{E}+\mathbf{v}\times\mathbf{B}].\label{forca}
$

In order to examine the effects of the noncommutative terms in Eq. \eqref{HAM} we first consider only configuration space NC, i.e, $\theta\neq 0$ and $\eta=0$:
\begin{align}
\mathbf{v}&={i\over\hbar}[H_{NC\theta},\mathbf{r}]={i\over\hbar}[c\boldsymbol{\alpha}\cdot\mathbf{p}+{e\over 2\hbar}[\nabla(c\boldsymbol{\alpha}\cdot\mathbf{A}-\Phi)\times\boldsymbol{\theta}]\cdot\mathbf{p}, \mathbf{r}] \nonumber \\ 
&=c\boldsymbol{\alpha}+{e\over 2\hbar}\nabla(c\boldsymbol{\alpha}\cdot\mathbf{A}-\Phi)\times\boldsymbol{\theta},
\end{align}
from which we conclude that the velocity of the particle \emph{is not gauge invariant} as the $\theta$-term in Eq. \eqref{HAM} breaks gauge symmetry explicitly. On these grounds we discard the $\theta$-term, taking $\theta=0$ in order to preserve gauge symmetry. Other ways to deal with this problem are discussed in Ref. \cite{Gitman}, where a suitable modification of the action is suggested. We do not pursue this proposal here.

Considering only momentum NC, i.e, $\eta\neq 0$ and $\theta=0$, the velocity is not modified:
\begin{align}
\mathbf{v}&={i\over\hbar}[H_{NC\eta},\mathbf{r}]={i\over\hbar}[c\boldsymbol{\alpha}\cdot\mathbf{p}+{c\over 2\hbar}[\boldsymbol{\alpha}\times\mathbf{r}]\cdot\boldsymbol{\eta}, \mathbf{r}]=c\boldsymbol{\alpha}.
\end{align}
Thus, momentum space NC preserves both gauge invariance and the noncommutative algebra, allowing for a natural generalization of the Dirac theory.

Rewriting the $\eta$-modified Hamiltonian as:
\begin{align}
H_{NC\eta}=c\boldsymbol{\alpha}\cdot(\mathbf{p}-{e\over 2}[\mathbf{B}+{1\over e\hbar}\boldsymbol{\eta}]\times\mathbf{r})+\beta mc^2 + e\Phi \label{etaH},
\end{align}
it is seen explicitly that the momentum NC consists in a shift of ${\boldsymbol{\eta}\over e\hbar}$ corresponding to a new electromagnetic potential given by:
\begin{equation}
\mathbf{A'}=\mathbf{A}+{\boldsymbol{\eta}\times\mathbf{r}\over 2e\hbar}\label{shift},
\end{equation}
as first pointed out in Ref. \cite{Nair}.

Since the electric field is not affected due to the time-independence of $\eta$, the equations of motion are only affected by:
\begin{align}
\mathbf{F}&={i\over\hbar}[H_\text{NC$\eta$},\boldsymbol{\pi}]+{\partial\boldsymbol{\pi}\over \partial t}=e[\mathbf{E}+\mathbf{v}\times(\mathbf{B}+{\boldsymbol{\eta}\over e\hbar})].
\end{align}

\subsection{Wave function}

The eigenfunctions of the new Hamiltonian, $\Psi_{NC}(x)$, are also modified; however, since the NC contribution corresponds to a constant shift in the vector \textbf{potential}, the NC correction yields a constant phase, which commutes with $H_\text{NC}$. Writing the new wave function as:
\begin{align}
\Psi_{NC}(x)=\exp\left\{{ie\over\hbar}\int_\text{traj.}\mathbf{A}'\cdot d\mathbf{l}\right\}\Psi(x)|_{\mathbf{A}=0} \label{phaseshift},
\end{align}
and hence, 
\begin{align}
\Psi_{NC}(x)=e^{{ie\over\hbar}\int_\text{traj.}(\mathbf{A}+{\boldsymbol{\eta}\times\mathbf{l}\over 2e\hbar})\cdot d\mathbf{l}}\Psi(x)|_{\mathbf{A}=0}=e^{{i\over 2\hbar^2}\int_\text{traj.}{\boldsymbol{\eta}\times\mathbf{l}}\cdot d\mathbf{l}}\Psi(x)\equiv e^{i\phi}\Psi(x) \label{ncphase}.
\end{align}
This result will be used in the next section in order to compute the hydrogen spectrum.

Notice that unless one encounters some topological obstruction or some crossing of energy levels one expects no Berry phase either \cite{OB4}.

\section{Hydrogen Atom}

The hydrogen atom can be solved in the non-relativistic limit, considering $e\Phi=-\alpha/r$, where $\alpha=e^2/ \hbar c$. Without a magnetic contribution, $\mathbf{A}=\mathbf{0}$, that is:
\begin{equation}
H_\text{NR hydrogen}=(c\boldsymbol{\alpha}\cdot\mathbf{p}+\beta m -{\alpha\over r}).
\end{equation}
The eigenvalues and eigenvectors of this Hamiltonian are well known and given by (from here onwards ones uses natural units, i.e. $\hbar=c=1$):
\begin{align}
E_n=-{m\alpha^2\over 2n^2},
\end{align}
where $n=1,2,...$ is the principal quantum number.

The observed complexity of the hydrogen spectrum can be understood by relativistic corrections, derived by perturbing the classical Hamiltonian using the \emph{Foldy-Wouthuysen transformation}. Then the Hamiltonian is decoupled into \cite{Bjorken}:
\begin{align}
H'''&=\beta\left(m+{(\mathbf{p}-e\mathbf{A})^2\over 2m}-{\mathbf{p}^4\over 8m^3}\right)+e\Phi-e{1\over 2m}\boldsymbol{\sigma}\cdot\mathbf{B}-{ie\over 8m^2}\boldsymbol{\sigma}\cdot\nabla\times\mathbf{E} \nonumber \\
&-{e\over 4m^2}\boldsymbol{\sigma}\cdot\mathbf{E}\times\mathbf{p}-{e\over8m^2}\nabla\cdot\mathbf{E}+\mathcal{O}(1/m^4).
\end{align}
The relativistic corrections can be categorized in terms of their dependence on the electric or magnetic field. The fine structure correction and 
Lamb shift are not affected by noncommutativity, but the \emph{hyperfine structure correction} is. This correction is derived from the expectation value of the interaction of the proton and electron magnetic moments, ${e\over2m}\boldsymbol{\sigma}\cdot\mathbf{B}$, breaking the $j$-degeneracy by splitting the energy of singlet and triplet states. 

Indeed, consider the perturbation:
\begin{align}
H_{hfs}=+{|e|\over2m}\boldsymbol{\sigma}\cdot\mathbf{B},
\label{hfs}
\end{align}
where $\mathbf{B}$ is the magnetic field acting on the electron, given by
\begin{align}
\mathbf{B}={g_pe\over2m_p}\int d^3r'\rho(r')\nabla\times(\boldsymbol{\sigma}_p\times\nabla){1\over4\pi|\mathbf{r}-\mathbf{r}'|}={g_pe\over3m_p}\boldsymbol{\sigma}_p \rho(r),
\end{align}
where $g_p$ is the gyromagnetic ratio, $\boldsymbol{\sigma}_p$ is the proton unit spin operator and $\rho(r)$, $m_p$ are the magnetic moment density and the mass of the proton, respectively. The resulting energy correction will be given by
\begin{align}
\Delta E_\text{hyperfine structure}=\left<\psi_{ln}|H_{hfs}|\psi_{ln}\right>=&{g_pe^2\over6mm_p}\left<\psi_{ln}|\boldsymbol{\sigma}\cdot\boldsymbol{\sigma}_p\rho(r)|\psi_{ln}\right> \nonumber \\
\approx&{g_pe^2\over6mm_p}(\boldsymbol{\sigma}\cdot\boldsymbol{\sigma}_p)|\psi_{ln}(0)|^2,
\end{align}
where the approximation $\rho(r)\approx\delta(r)$ was made. In the case of states with $l=0$, $|\psi_{0n}(0)|^2=4m^3\alpha^4/n^3e^2$, and the eigenvalues of the spin operator $\boldsymbol{\sigma}\cdot\boldsymbol{\sigma}_p$ are 1/2 for triplet states (the spins are aligned) and $-3/2$ for singlet states (the spins are oppositely aligned). 
Therefore, the energy split between aligned and anti-aligned $s$-states will be 
\begin{align}
\delta_n={4\over3}{g_p\alpha^4m^2\over n^3m_p}.
\end{align}
In particular, for the $1S_{1/2}$ state, the energy splitting is given by:
\begin{align}
\delta_{1S_{1/2}}={4\over 3}{g_p\alpha^4m^2\over m_p}\approx 5.88~\mu eV,
\end{align}
corresponding to a transition that emits a photon with a frequency of approximately 1420 MHz. Actually, the experimental accuracy of this result is remarkable, and in fact it is regarded as an established value. Its measurement yields $1420405751.7691\pm0.0024$Hz \cite{1S1970}.

\subsection{The noncommutative hydrogen atom}

The previous calculations can be adapted to the noncommutative case, since the noncommutativity affects only the hyperfine structure.

Notice that the non-relativistic treatment of the hydrogen atom does not involve the magnetic field, so the non-relativistic energy levels are unaltered. 
Furthermore, as seen before, the NC wave function will be modified just by a phase at all orders, and therefore also in the relativistic limit.

The noncommutative correction to the hyperfine term is given by:
\begin{align}
H_{NChfs}=+{|e|\over2m}\boldsymbol{\sigma}\cdot(\mathbf{B}+{\boldsymbol{\eta}\over e})=H_{hfs}+{1\over2m}\boldsymbol{\sigma}\cdot\boldsymbol{\eta}.
\end{align}
Hence, the expectation value of this perturbation for the eigenstates of the perturbed noncommutative Hamiltonian $H_{NC}$, $\psi'_n$, is:
\begin{align}
\Delta E_\text{hyperfine structure}^{NC}=&\left<\psi'_n|H_{NChfs}|\psi'_n\right> \nonumber \\ 
=&\left<e^{i\phi}\psi_n|H_{NChfs}|e^{i\phi}\psi_n\right> \nonumber \\
=&\left<\psi_n|H_{hfs}|\psi_n\right> +{1\over2m} \left<\psi_n|\boldsymbol{\sigma}\cdot\boldsymbol{\eta}|\psi_n\right> \nonumber \\
=&\Delta E_\text{hyperfine structure} +{1\over2m} \left<\psi_n|\boldsymbol{\sigma}\cdot\boldsymbol{\eta}|\psi_n\right>,
\end{align}
where $\boldsymbol{\eta}=\eta\mathbf{e}$, and $\mathbf{e}$ is a unit vector in an arbitrary direction (it only depends on the definition of the antisymmetric matrix $\epsilon_{ij}$). If $\mathbf{e}$ is aligned with the proton spin ($\mathbf{e}=\boldsymbol{\sigma}_p$), the eigenvalues of $\boldsymbol{\sigma}\cdot\boldsymbol{\sigma}_p$ coincide with the eigenvalues of $\boldsymbol{\sigma}\cdot\mathbf{e}$, 1/2 for singlet states and -3/2 for triplets.

Using the orthonormality of the states $\psi_n$, we find the noncommutative energy difference between up and down states will be independent of energy level, $n$:
\begin{align}
\delta_{NC}={\eta\over 2m}\epsilon\cos\phi ,
\end{align} 
where $\phi$ is the angle between $\boldsymbol{\sigma}_p$ and $\boldsymbol{e}$, such that $\boldsymbol{\sigma}_p\cdot\boldsymbol{e}=\epsilon\cos\phi$. It follows that the modified hyperfine transition energy is shifted by:
\begin{align}
\delta_{n}^{NC}=\delta_n+ {\eta\over 2m}\epsilon\cos\phi.
\end{align}
Since the value $h^{-1}\delta_{n}$ is known with an accuracy of $0.0024~Hz$ for the $1S_{1/2}$ state, it can be used to set an upper bound to the value of $\eta$, as ${1\over2}\epsilon\cos\phi=\mathcal{O}(1)$:
\begin{align}
&\eta\lsim m h\times0.0024=1.45\times10^{-66}{~Kg^2m^2s^{-2}} \nonumber \\
\Rightarrow&\sqrt{\eta}\lsim 2.26~\mu eV/c.
\end{align}

This is the main result of this work, and constitutes in a quite stringent bound on the momentum noncommutative parameter. It is as tight as the bound obtained from the spectrum of the gravitation quantum well \cite{OB1}, even though that bound was obtained through an assumption for the configuration space noncommutative parameter, $\theta$, namely that $\sqrt{\theta}\lesssim1TeV$. Notice that the gravitational quantum well yields a bound for the product $\theta\eta/4\hbar^2\lesssim10^{-24}$, which can be directly compared with the one arising from entropic gravity and the Equivalence Principle considerations, namely $\theta\eta/\hbar^2\lesssim10^{-13}$ \cite{BBDP2011}, as it involves the same combination of noncommutative parameters.

\section{Conclusions}

In this work phase space noncommutativity effects on the Dirac equation and its impact on the dynamics of a spin-1/2 particle are studied. This allows for acquiring some new insight on the nature of the noncommutative effects.

It has been shown that, without changing the original Hamiltonian, in order to preserve gauge invariance, one should restrict the analysis to the case where the coordinates commute, i.e., $\theta=0$, and consider only momentum noncommutativity. 
The introduction of momentum non-commutativity implies the appearance of a constant magnetic field-like term in the 
Hamiltonian, and it is shown that the resulting shift in the spectrum of the hydrogen atom has an observational signature.

The effect of momentum NC in the spectrum of the hydrogen atom is analyzed in detail and shown that it shifts the hyperfine structure split. From the available experimental accuracy for the hyperfine structure energy, one is able to set an upper bound $\sqrt{\eta}\lsim 2.26\mu eV/c$, which is consistent with other values obtained in the literature \cite{OB1} and indicates that noncommutative effects are fairly small, meaning that if relevant, they may arise in high energy phenomena. Indeed, recent studies reveal that momentum NC effects have a considerable impact in quantum cosmology \cite{9} and black hole physics \cite{10, 11, 12}.

\vskip 0.5cm

\noindent
{\bf Acknowlegments}

\vskip 0.5cm

\noindent
The authors would like to thank Carlos Zarro for interesting comments and suggestions.


\vskip 3cm


\begin{thebibliography}{99}        



\bibitem{DS} M.R. Douglas, N.A. Nekrasov, Rev. Mod. Phys. {\bf 73} (2001) 977; R. Szabo, Phys. Rep. {\bf 378} (2003) 207.

\bibitem{Carroll} S.M. Carroll, J.A. Harvey, V.A. Kosteleck\'y, C.D. Lane, T. Okamoto, Phys. Rev. Lett. \textbf{87} (2001) 141601.

\bibitem{BGuisado2} O. Bertolami and L. Guisado, JHEP {\bf 0312} (2003) 013.

\bibitem{CS} A. Connes, M.R. Douglas and A. Schwarz, JHEP {\bf 9802} (1998) 003; N. Seiberg and E. Witten, JHEP {\bf 9909} (1999) 032.

\bibitem{4} J. Gamboa, M. Loewe and J. C. Rojas, Phys. Rev. {\bf D64} (2001) 067901.

\bibitem{Nair} V. P. Nair and A. P. Polychronakos, Phys. Lett. \textbf{B505} (2001) 267.

\bibitem{Duval} C. Duval and P. A. Horvathy, Journ. Phys. {\bf A34} (2001) 10097.

\bibitem{Ho} Pei-Ming Ho and Hsien-Chung Kao, Phys. Rev. Lett. \textbf{88} (2002) 151602.

\bibitem{Horvathy} P. A. Horvathy, Ann. Phys. (N. Y.) {\bf 299} (2002) 128.

\bibitem{5} Jian-zu Zhang, Phys. Rev. Lett {\bf 93} (2004) 043002; Phys. Lett. {\bf B584} (2004) 204.

\bibitem{OB1} O. Bertolami, J. G. Rosa, C. M. L. Arag\~ao, P. Castorina and D. Zappal\`a, Phys. Rev. {\bf D72} (2005) 025010.

\bibitem{7} C. Acatrinei, Mod. Phys. Lett. {\bf A20} (2005) 1437.

\bibitem{8} O. Bertolami, J. G. Rosa, C. M. L. Arag\~ao, P. Castorina and D. Zappal\`a, Mod. Phys. Lett. {\bf A21} (2006) 795.

\bibitem{OB3} C. Bastos, O. Bertolami, N. C. Dias and J. N. Prata, J. Math. Phys. {\bf 49} (2008) 072101. 

\bibitem{9} C. Bastos, O. Bertolami, N. C. Dias and J. N. Prata, Phys. Rev. {\bf D78} (2008) 023516. 

\bibitem{10} C. Bastos, O. Bertolami, N. C. Dias and J. N. Prata, Phys. Rev. {\bf D80} (2009) 124038. 

\bibitem{11} C. Bastos, O. Bertolami, N. C. Dias and J. N. Prata, Phys. Rev. {\bf D82} (2010) 041502. 

\bibitem{12} C. Bastos, O. Bertolami, N. C. Dias and J. N. Prata, Phys. Rev. {\bf D84} (2011) 024005.

\bibitem{Gitman} T. C. Adorno, M. C. Baldiotti and D. M. Gitman, arXiv:1008.4890v2 [hep-th] (2010).

\bibitem{OB4} C. Bastos and O. Bertolami,  Phys. Lett. {\bf A372} (2008) 5556.

\bibitem{Bjorken} J. D. Bjorken and S. D. Drell, {\it Relativistic Quantum Mechanics}, McGraw-Hill Book Company (1964).

\bibitem{1S1970} H. Hellwig, R. F. C. Vessot, M. W. Levine, P. W. Zitzewitz, D. W. Allan and D. J. Glaze, IEEE Trans. Instrum. Meas. {\bf IM-19} (1970), 200.  

\bibitem{BBDP2011} C. Bastos, O. Bertolami, N. C. Dias and J. N. Prata, Class. Quantum Grav. {\bf 28} (2011) 125007.


\[
\]%
\[
\]

\end{thebibliography}
\end{document}